\documentclass[11pt]{article} 
\usepackage{amssymb,amsmath,amscd,xspace,amsthm,tocvsec2,mathrsfs,color}
\newcommand{\be}{\begin{equation}}
\newcommand{\ee}{\end{equation}}
\newcommand{\bea}{\begin{eqnarray}}
\newcommand{\eea}{\end{eqnarray}}
\newcommand{\sn}{{\rm sn}}

\newcommand{\dn}{{\rm dn}}
\newcommand{\cn}{{\rm cn}}
\newcommand{\sech}{{\rm sech}}
\newcommand{\R}{\mathbb R}

\theoremstyle{plain}
\newtheorem{theorem}{Theorem}

\theoremstyle{definition}

\begin{document}
\vspace{.5in} 
\begin{center} 
{\LARGE{\bf Solutions of Several Coupled Discrete Models 
in terms of Lam\'e Polynomials of Arbitrary Order}}
\end{center} 

\vspace{.3in}
\begin{center} 
{\LARGE{\bf Avinash Khare}} \\ 
{Raja Ramanna Fellow, Indian Institute of Science Education and Research,
Pune 411021, India}
\end{center} 

\begin{center} 
{\LARGE{\bf Avadh Saxena}} \\ 
{Theoretical Division and Center for Nonlinear Studies, Los Alamos
National Laboratory, Los Alamos, NM 87545, USA}
\end{center} 

\begin{center}
{\LARGE{\bf Apoorva Khare}} \\
{Departments of Mathematics and Statistics, Stanford University,
CA 94305, USA}
\end{center}

\vspace{.9in}
{\bf {Abstract:}}  

Coupled discrete models abound in several areas of physics. Here we
provide an extensive set of exact quasiperiodic solutions of a number of
coupled discrete models in terms of Lam\'e polynomials of arbitrary
order. The models discussed are (i) coupled Salerno model, (ii) coupled
Ablowitz-Ladik model, (iii) coupled $\phi^4$  model, and (iv) coupled
$\phi^6$ model. In all these cases we show that the coefficients of the
Lam\'e polynomials are such that the Lam\'e polynomials can be
reexpressed in terms of Chebyshev polynomials of the relevant Jacobi
elliptic function.\medskip

\newpage 
  
\section{Introduction} 
In a recent paper \cite{ak11} we have obtained solutions of a number of 
coupled discrete models in terms of Lam\'e polynomials of order one and
two. The purpose of the present paper is 
to show that in the same models one can in fact obtain solutions in terms
of Lam\'e polynomials (and hence trigonometric and hyperbolic polynomials) 
of arbitrary order. 
In particular, we obtain solutions of (i) Coupled Salerno
model (ii) coupled Ablowitz-Ladik (AL) model (iii) coupled $\phi^6$ model
(iv) coupled $\phi^4$ model, in terms of Lam\'e (and hence trigonometric
and hyperbolic) polynomials of arbitrary order.
As an illustration, we confine our discussion to coupled Salerno model
and then show how similar solutions also exist in coupled 
Ablowitz-Ladik, coupled $\phi^4$ and coupled $\phi^6$ models. 
Quite remarkably, we find that the coefficients of the Lam\'e polynomials are
such that the Lame polynomials can be reexpressed as
Chebyshev polynomials of the relevant Jacobi elliptic function. 

There are many physical situations where a discrete field theory is appropriate 
to model the phenomena of interest with a specific coupling between the two
fields.   This fact serves as one of the main motivation for the results presented 
here.  A technologically important phenomenon of current intense interest is the
coexistence of magnetism and ferroelectricity (i.e., magnetoelectricity)
in a given material. This is a highly desired functionality in applications
involving cross-field response such as electric field control of magnetism and 
magnetic field control of electric polarization, switching and actuation.  More 
generally, presence of two or more such ferroic properties (e.g. polarization, 
magnetization, strain) is referred to as multiferroic behavior \cite{multif}.  In 
recent years, two different classes of (single phase) multiferroics, namely the 
orthorhombically distorted perovskites \cite{kimura} and rare earth hexagonal 
structures \cite{fiebig}, have emerged in addition to other related structures.  
The hexagonal materials show axial polarization and magnetic domain walls 
in the basal planes.  The latter can be modeled by a coupled $\phi^4$ model 
\cite{curnoe} in the presence of a magnetic field.   Coupled $\phi^4$ models 
\cite{aubry,abel,jcp} are useful in the study of many ferroelectric and other 
second order phase transitions.  For multiferroic materials the relevant coupled 
$\phi^4$ model \cite{curnoe} has a biquadratic coupling.  In contrast, the coupled 
$\phi^4$ model for a surface phase transition with hydration forces \cite{jcp}, 
which arises in the biophysics context, has a bilinear coupling.  Other types of 
couplings are also known for structural phase transitions involving strain  tensor 
components \cite{das}.  

There are many examples of coupled discrete Ablowitz-Ladik, coupled discrete 
Salerno and coupled saturated discrete nonlinear Schr\"odinger (DNLS) models 
known in the literature \cite{cuevas,rothos,panosbook}.  In addition, there are 
analogous coupled models known in the field theoretic contexts \cite{raja,pla}.  
Several related models have been discussed in the literature and their soliton 
solutions have been obtained \cite{lai,rao,wang,huang,bazeia,zhu,lou,cao} 
including periodic ones \cite{li,liu,llw}.  

The paper is organized as follows. 
In Section II we provide the solutions for the coupled Salerno model in
terms of Lam\'e polynomials of order three and four. 
Based on these results as well those obtained in \cite{ak11}, in Section III 
we generalize these results and conjecture
solutions in terms of Lam\'e (and hence trigonometric and hyperbolic) 
polynomials of arbitrary order. In Section
IV we show that these Lam\'e polynomials can be reexpressed as Chebyshev
polynomials of the relevant Jacobi elliptic function. 
This also proves that our proposed Lam\'e polynomials of arbitrary
order are indeed solutions of the coupled equations.
In Section V we show how the coupled Ablowitz-Ladik, coupled discrete
$\phi^6$, and coupled $\phi^4$ models also admit solutions in terms of
Lam\'e polynomials of arbitrary order.
Section VI contains the summary of main results and possible future
directions.

\section{Coupled Salerno Model}
 
As discussed in a recent paper \cite{ak11}, the field equations of the 
coupled Salerno model are given by
\be\label{m8}
idu_n/dt +[u_{n+1}+u_{n-1}-2u_n]+(\mu_1 \mid u_n \mid^2+\mu_2\mid
v_n \mid^2)\left[u_{n+1}+u_{n-1}+\frac{\nu_1-2\mu_1}{\mu_1}u_n\right] =0\,,
\ee
\be\label{m9}
idv_n/dt +\left[v_{n+1}+v_{n-1}-\left(2+\frac{\nu_1 \mu_2}{\mu_1^2}
-\frac{\nu_2}{\mu_2}\right)v_n\right]+(\mu_1 \mid u_n \mid^2+\mu_2\mid
v_n \mid^2)\left[v_{n+1}+v_{n-1}+\frac{\nu_2-2\mu_2}{\mu_2}v_n\right] =0\,.
\ee
In (\cite{ak11}) we have already obtained solutions of these coupled equations
in terms of Lam\'e polynomials of order one and two. We now show that the same
model also admits Lam\'e polynomial solutions of arbitrary order. 

As in \cite{ak11}, we start with the ansatz
\be\label{m10}
u_n=f_n \exp{[-i(\omega_1 t+\delta_1)]}\,,
~~v_n=g_n \exp{[-i(\omega_2 t+\delta_2)]}\,,
\ee
then it is easily shown that the above coupled equations take the form
\be\label{m11}
(\omega_1-2) f_n +(f_{n+1}+f_{n-1})(\mu_1 f_n^2+\mu_2 g_n^2 +1)
+(\mu_1 f_n^2+ \mu_2 g_n^2)\frac{(\nu_1-2\mu_1)}{\mu_1}f_n =0\,,
\ee
\be\label{m12}
\left(\omega_2 - 2+\frac{\nu_1 \mu_2}{\mu_1^2}-\frac{\nu_2}{\mu_2}\right) g_n 
+(\mu_1 f_n^2+\mu_2 g_n^2 +1) (g_{n+1}+g_{n-1})
+(\mu_1 f_n^2+ \mu_2 g_n^2) \frac{(\nu_2-2\mu_2)}{\mu_2}g_n =0\,.
\ee
It is clear from Eqs. (\ref{m11}) and (\ref{m12}) that in general these
coupled equations will have exact solutions if
\be\label{m13}
1+\mu_1 f_n^2 + \mu_2 g_n^2 = 0\,,
\ee
and in that case
\be\label{m14}
\omega_1 = \frac{\nu_1}{\mu_1}\,,~~\omega_2 = \frac{\nu_1 \mu_2}{\mu_1^2}\,.
\ee
We now show that Eq. (\ref{m13}) has solutions in terms of Lam\'e
polynomials of arbitrary order. In particular, we show that at every
order there are three distinct solutions satisfying Eq. (\ref{m13}). As a
first step, let us explicitly obtain solutions of Eq. (\ref{m13}) in
terms of Lam\'e polynomials of order three and four.  
(It may be noted that we have already obtained such solutions before in
terms of Lame polynomials of order one and two \cite{ak11}.)

\subsection{Lam\'e Polynomial Solutions of Order Three}

It is easily checked that one of the exact solution to Eq. (\ref{m13}) is 
\be\label{m15}
f_n= \dn [\beta(n+c_2),m] (A \dn^2 [\beta(n+c_2),m]+B) \,,
~~ g_n= \sqrt{m} \sn[\beta(n+c_2),m] (C \dn^2 [\beta(n+c_2),m]+D) \,,
\ee
provided
\bea\label{m16}
&&\mu_1,\mu_2<0\,,~~|\mu_1|A^2=|\mu_2|C^2\,,~~|\mu_2|D^2=1\,, \nonumber \\
&&|\mu_1|(A+B)^2 = 1\,,~~|\mu_1|(A^2+2AB)=2CD|\mu_2|\,.
\eea
Here $c_2$ is an arbitrary constant signifying discrete translation
invariance. Also notice that for this solution, the width $\beta$ is also
completely arbitrary. We shall see that all the solutions discussed in
this paper are valid for arbitrary $c_2$ and width $\beta$. On solving,
we find that 
\be\label{m17}
\sqrt{|\mu_1|}A = 4\,,B=-\frac{3}{4}A\,,~~\sqrt{|\mu_2|}C =4\,,~~C=-4D\,.
\ee
Since the field Eqs. (\ref{m11}) and (\ref{m12}) are invariant under $f_n
\rightarrow \pm f_n,g_n \rightarrow \pm g_n$, one can trivially
write three other solutions from here.  
Note that $\beta$ is completely arbitrary. Using the fact that $\dn(x,m)$
has period $2K(m)$ while $\cn(x,m)$ and $\sn(x,m)$ are periodic functions
with period $4K(m)$, it follows that for the solution (\ref{m15}),
$f_n,g_n$ and hence $u_n,v_n$ satisfy the boundary condition
\be\label{m17a}
u_{n+\frac{2K(m)}{\beta}}= u_n\,,~~v_{n+\frac{4K(m)}{\beta}}=v_n\,.
\ee
Here $K(m)$ is the complete elliptic integral of the first kind.

Another solution to Eq. (\ref{m13}) is 
\be\label{m18}
f_n= \cn [\beta(n+c_2),m] (A \cn^2 [\beta(n+c_2),m]+B) \,,
~~ g_n= \sn[\beta(n+c_2),m] (C \cn^2 [\beta(n+c_2),m]+D) \,,
\ee
provided $A,B,C,D$ are again given by Eq. (\ref{m17}). Note however that
the solution (\ref{m18}) is distinct from the solution (\ref{m15}). In
particular, while for the solution (\ref{m15}), $u_n,v_n$ satisfy the
boundary condition (\ref{m17a}), for the solution (\ref{m18}), $f_n,g_n$
and hence $u_n,v_n$ satisfy the boundary condition
\be\label{m19}
u_{n+\frac{2K(m)}{\beta}}= u_n\,,~~v_{n+\frac{2K(m)}{\beta}}=v_n\,.
\ee

However, in the limit $m=1$, both the solutions (\ref{m15}) and
(\ref{m18}) go over to the hyperbolic solution
\be\label{m20}
f_n=  A \sech^3 [\beta(n+c_2)]+B \sech [\beta(n+c_2)] \,,
~~ g_n= \tanh [\beta(n+c_2)] (C \sech^2 [\beta(n+c_2)]+D) \,.
\ee

Further, in the limit $m=0$, the solution (\ref{m18}) goes over to the
trigonometric solution
\be\label{m18t}
f_n= \cos [\beta(n+c_2)] (A \cos^2 [\beta(n+c_2)]+B) \,,
~~ g_n= \sin[\beta(n+c_2)] (C \cos^2 [\beta(n+c_2)]+D) \,.
\ee
All the four solutions discussed above are valid only if $\mu_1,\mu_2<0$. 
If instead $\mu_1,\mu_2$ have opposite signs then also one has solutions to 
Eq. (\ref{m13}). One such solution is
\be\label{m21}
f_n= \frac{1}{\dn [\beta(n+c_2),m]} \left(\frac{A} {\dn^2 [\beta(n+c_2),m]}+B\right) \,,
~~ g_n= \frac{\sqrt{m} \sn[\beta(n+c_2),m]}
{\dn [\beta(n+c_2),m]} \left(\frac{C}{\dn^2 [\beta(n+c_2),m]}+D\right) \,,
\ee
provided $\mu_1<0,\mu_2>0$ while $A,B,C,D$ are still given by Eq. (\ref{m17}).
Note that for this solution $f_n,g_n$
and hence $u_n,v_n$ satisfy the boundary condition (\ref{m17a}). Further, 
if we interchange $f_n$ and $g_n$, then $\mu_1>0,\mu_2<0$. In the limit $m=1$,
this solution goes over to the hyperbolic solution
\be\label{m21a}
f_n=  A \cosh^3 [\beta(n+c_2)]+B \cosh [\beta(n+c_2)] \,,
~~ g_n= \sinh [\beta(n+c_2)] (C \cosh^2 [\beta(n+c_2)]+D) \,.
\ee

\subsection{Lam\'e Polynomial Solutions of Order Four}

It is easily checked that one of the exact solution to Eq. (\ref{m13}) is 
\bea\label{m22}
&&f_n= A \dn^4 [\beta(n+c_2),m] +B \dn^2 [\beta(n+c_2),m] +C\,, \nonumber \\
&&g_n= \sqrt{m} \sn[\beta(n+c_2),m] \dn [\beta(n+c_2),m] 
(D \dn^2 [\beta(n+c_2),m] + E) \,,
\eea
provided
\bea\label{m23}
&&\mu_1,\mu_2<0\,,~~|\mu_1|A^2=|\mu_2|D^2\,,~~|\mu_1|C^2=1\,, 
~~A(A+2B)|\mu_1|= 2DE|\mu_2|\,, \nonumber \\
&&|\mu_1|(A+B)^2 +2AC|\mu_1|= |\mu_2|E^2\,,~~2BC|\mu_1|=-|\mu_2|E^2\,.
\eea
On solving, we find that 
\be\label{m24}
\sqrt{|\mu_1|}C = 1\,,A=-B=8C\,,~~\sqrt{|\mu_2|}E = -4\,,~~D=-2E\,.
\ee
For the solution (\ref{m22}), 
$f_n,g_n$ and hence $u_n,v_n$ satisfy the boundary condition (\ref{m17a}).

Another solution to Eq. (\ref{m13}) is 
\bea\label{m25}
&&f_n= A \cn^4 [\beta(n+c_2),m]+B \cn^2 [\beta(n+c_2),m]+ C\,, \nonumber \\
&&g_n= \sn[\beta(n+c_2),m] \cn[\beta(n+c_2),m] 
(D \cn^2 [\beta(n+c_2),m]+E)\,,
\eea
provided $A,B,C,D,E$ are again given by Eq. (\ref{m24}). Note however that
the solution (\ref{m25}) is distinct from the solution (\ref{m22}). In
particular, while for the solution (\ref{m22}), $u_n,v_n$ satisfy the
boundary condition (\ref{m17a}), for the solution (\ref{m25}), $f_n,g_n$
and hence $u_n,v_n$ satisfy the boundary condition (\ref{m19}).

However, in the limit $m=1$, both the solutions (\ref{m22}) and (\ref{m25}) go 
over to the hyperbolic solution
\bea\label{m26}
&&f_n=  A \sech^4 [\beta(n+c_2)]+B \sech^2 [\beta(n+c_2)] +C\,, \nonumber \\
&&g_n= \tanh [\beta(n+c_2)] \sech [\beta(n+c_2)] 
(D \sech^2 [\beta(n+c_2)]+E) \,,
\eea

Further, in the limit $m=0$, the solution (\ref{m25}) goes over to the
trigonometric solution
\bea\label{m25t}
&&f_n= A \cos^4 [\beta(n+c_2)]+B \cos^2 [\beta(n+c_2)]+ C\,, \nonumber \\
&&g_n= \sin[\beta(n+c_2)] \cos[\beta(n+c_2)] 
(D \cos^2 [\beta(n+c_2)]+E)\,,
\eea

All the four solutions discussed above are valid only if
$\mu_1,\mu_2<0$. If instead $\mu_1,\mu_2$ have opposite signs then also
one has solutions to Eq. (\ref{m13}). One such solution is
\bea\label{m27}
&&f_n= \frac{A}{\dn^4 [\beta(n+c_2),m]} 
+\frac{B} {\dn^2 [\beta(n+c_2),m]}+C \,, \nonumber \\
&&g_n= \sqrt{m} \sn[\beta(n+c_2),m]
\left(\frac{D}{\dn^4 [\beta(n+c_2),m]} +\frac{E}{\dn^2 [\beta(n+c_2),m]}\right) \,,
\eea
provided $\mu_1<0,\mu_2>0$ while $A,B,C,D,E$ are still given by Eq.
(\ref{m24}). For this solution, $f_n,g_n$ and hence $u_n,v_n$ satisfy the
boundary condition (\ref{m17a}). Note that if we interchange $f_n$ and
$g_n$, then $\mu_1>0,\mu_2<0$. In the limit $m=1$, this solution goes
over to the hyperbolic solution
\bea\label{m28}
&&f_n=  A \cosh^4 [\beta(n+c_2)]+B \cosh^2 [\beta(n+c_2)]+C\,, \nonumber \\
&&g_n= \sinh [\beta(n+c_2)] (D \cosh^3 [\beta(n+c_2)]
+E \cosh[\beta(n+c_2)]) \,.
\eea

\section{General Results}

Looking at the structure of the solutions in terms of Lam\'e polynomials
of order one to four, it is easy to generalize and write down the
solutions of Eq. (\ref{m13}) in terms of Lam\'e polynomials of arbitrary
order. 
For this we need to divide the discussion into two parts depending on if
we are considering Lam\'e polynomials of odd or even order.

{\bf Case I: Lame Polynomials of odd order}

One of the solutions can be written in the form (note $n$ is an odd
integer)
\bea\label{n1}
&&f_n = \sum_{k=1}^{(n+1)/2} A_k (\dn[\beta(n+c_2),m])^{2k-1}\,,
\nonumber \\  
&&g_n = \sqrt{m} \sn[\beta(n+c_2),m] \sum_{k=1}^{(n+1)/2} B_k 
(\dn[\beta(n+c_2),m])^{2k-2}\,.
\eea
For this solution, $f_n,g_n$ and hence $u_n,v_n$ satisfy the boundary
condition (\ref{m17a}). 
Note that there are $(n+1)/2$ number of terms in both $f_n$ and $g_n$. 

Another solution is given by
\bea\label{n1a}
&&f_n = \sum_{k=1}^{(n+1)/2} A_k (\cn[\beta(n+c_2),m])^{2k-1}\,,
\nonumber \\  
&&g_n = \sn[\beta(n+c_2),m] \sum_{k=1}^{(n+1)/2} B_k 
(\cn[\beta(n+c_2),m])^{2k-2}\,.
\eea
For this solution, $f_n,g_n$ and hence $u_n,v_n$ satisfy the boundary
condition (\ref{m19}). 

In the limit $m=1$, both the solutions (\ref{n1}) and (\ref{n1a}) 
go over to the hyperbolic solution
\bea\label{n1b}
&&f_n = \sum_{k=1}^{(n+1)/2} A_k (\sech [\beta(n+c_2)])^{2k-1}\,,
\nonumber \\  
&&g_n = \tanh[\beta(n+c_2)] \sum_{k=1}^{(n+1)/2} B_k 
(\sech [\beta(n+c_2)])^{2k-2}\,.
\eea

Further, in the limit $m=0$, the solution (\ref{n1a}) goes over to the
trigonometric solution
\bea\label{n1aa}
&&f_n = \sum_{k=1}^{(n+1)/2} A_k (\cos[\beta(n+c_2)])^{2k-1}\,,
\nonumber \\  
&&g_n = \sin[\beta(n+c_2)] \sum_{k=1}^{(n+1)/2} B_k 
(\cos[\beta(n+c_2)])^{2k-2}\,.
\eea

Note that these four solutions are valid in case $\mu_1,\mu_2 <0$.  
However, if $\mu_1$ and $\mu_2$ have opposite signs, say $\mu_1 <0$, 
$\mu_2 >0$, then the solution is given by
\bea\label{n1c}
&&f_n = \sum_{k=1}^{(n+1)/2} \frac{A_k}{(\dn[\beta(n+c_2),m])^{2k-1}}\,,
\nonumber \\  
&&g_n = \sqrt{m} \sn[\beta(n+c_2),m] \sum_{k=1}^{(n+1)/2} 
\frac{B_k}{(\dn[\beta(n+c_2),m])^{2k-1}}\,.
\eea
For this solution, $f_n,g_n$ and hence $u_n,v_n$ satisfy the boundary
condition (\ref{m17a}). In the limit $m=1$, this solution goes over to
the hyperbolic solution  
\bea\label{n1d}
&&f_n = \sum_{k=1}^{(n+1)/2} A_k (\cosh [\beta(n+c_2)])^{2k-1}\,,
\nonumber \\  
&&g_n = \tanh[\beta(n+c_2)] \sum_{k=1}^{(n+1)/2} B_k 
(\cosh [\beta(n+c_2)])^{2k-1}\,.
\eea

On substituting any of the expressions for $f_n,g_n$ as given by Eqs.
(\ref{n1}) to (\ref{n1d}) in Eq. (\ref{m13}), we obtain $n+1$ equations
which determine the $n+1$ parameters $A_k,B_k$. While all $A_k$'s are
numbers in units of $\frac{1}{\sqrt{|\mu_1|}}$, all $B_k$'s are numbers
in units of $\frac{1}{\sqrt{|\mu_2|}}$. For simplicity from now onwards
we will merely give the numerical values of $A_k,B_k$ and it is
understood that they are in units of $\frac{1}{\sqrt{|\mu_1|}}$ and
$\frac{1}{\sqrt{|\mu_2|}}$, respectively. 
Some of the relations are
\bea\label{n2}
&&A_{(n+1)/2}^2 = B_{(n+1)/2}^2\,, \nonumber \\
&&2A_{(n+1)/2} A_{(n-1)/2}+B_{(n+1)/2}^2 = 2B_{(n+1)/2} B_{(n-1)/2}\,, 
\nonumber \\
&&A_{(n-1)/2}^2+2A_{(n+1)/2} A_{(n-3)/2}+2B_{(n+1)/2} B_{(n-1)/2}= 
B_{(n-1)/2}^2 + 2B_{(n+1)/2} B_{(n-3)/2}\,, \nonumber \\ 
&&B_{(n-1)/2}^2+2A_{(n+1)/2} A_{(n-5)/2}+2B_{(n+1)/2} B_{(n-3)/2}
+2A_{(n-1)/2}A_{(n-3)/2} \nonumber \\
&&= 2 B_{(n-1)/2} B_{(n-3)/2}+ 2B_{(n+1)/2} B_{(n-5)/2}\,, \nonumber \\ 
&&B_1^2 = 1\,,~~A_1^2+2B_1 B_2 = B_1^2\,,~~
2A_1 A_2+B_2^2+2B_1 B_3 = 2B_1 B_2\,, \nonumber \\
&&A_2^2+2A_1 A_3-B_2^2+2B_2 B_3+2B_1 B_4- 2B_1 B_3 = 0\,. 
\eea
On comparing these results with the exact expressions for $n=1,3$,
we conjecture the following general results for arbitrary odd $n$:
\bea\label{n3}
&&A_{(n+1)/2}=B_{(n+1)/2} = 2^{n-1}\,,
~~A_{(n-1)/2}=-n 2^{n-3}\,,~~B_{(n-1)/2}=-(n-2)2^{n-3}\,, \nonumber \\
&&A_{(n-3)/2}=\frac{n(n-3)}{2!} 2^{n-5}\,,
~~B_{(n-3)/2}=\frac{(n-3)(n-4)}{2!} 2^{n-5}\,,
\nonumber \\
&&A_{(n-5)/2}=-\frac{n(n-4)(n-5)}{3!} 2^{n-7}\,,
~~B_{(n-5)/2}=-\frac{(n-4)(n-5)(n-6)}{3!}  2^{n-7}\,, \nonumber \\
&&\sum_{i=1}^{(n+1)/2} A_i = 1\,,~~
\sum_{i=1}^{(n+1)/2} B_i = n\,.
\eea
Further, depending on if $n=4k+3$ or $4k+1$ we have the following results:
\be\label{n4}
A_1=-n\,, ~~~B_1=-1\,, ~~if~~n=4k+3\,,~~k=0,1,2,...
\ee
\be\label{n5}
A_1=+n\,, ~~~B_1=+1\,, ~~if~~n=4k+1\,,~~k=0,1,2,...\,. 
\ee

In fact looking at these general results, we conjecture the following
expressions for various $A_i,B_i$.
\be\label{g1}
A_{\frac{n-(2k-1)}{2}} = (-1)^{k} \frac{n(n-k-1)(n-k-2)...(n-2k+1)}
{k!} 2^{n-2k-1}\,,~~k=0,2,...,\frac{n-1}{2}\,,
\ee
\be\label{g2}
B_{\frac{n-(2k-1)}{2}} = (-1)^{k} \frac{(n-k-1)(n-k-2)...(n-2k)}
{k!} 2^{n-2k-1}\,,~~k=0,1,2,...,\frac{n-1}{2}\,.
\ee

It can be verified that for $n=1,3$, the $A_i, B_i$ which follow
from Eqs. (\ref{n3}) to (\ref{n5}) 
agree with the values obtained by us above as well as in \cite{ak11}.
Further, for low values of (odd) $n$, it can be numerically checked that
the $A_i, B_i$ which follow from Eqs. (\ref{g1}) and (\ref{g2}) indeed
satisfy the various relations which follow by demanding the validity of
the constraint relation (\ref{m13}).

In the next section, we will prove that our conjectures hold for all
$n$. But first, using Eqs. (\ref{n3}), (\ref{g1}), and (\ref{g2}), we
predict that the following identities hold for all odd $n$:
\be\label{g5}
\sum_{k=0}^{(n-1)/2}  (-1)^{k} \frac{n(n-k-1)(n-k-2)...(n-2k+1)}
{k!} 2^{n-2k-1} =1\,.
\ee
\be\label{g6}
\sum_{k=0}^{(n-1)/2}  (-1)^{k} \frac{(n-k-1)(n-k-2)...(n-2k)}
{k!} 2^{n-2k-1} = n\,.
\ee
We will see that these identities naturally lead to the
proof of our conjectures.

{\bf Case II: Lame Polynomials of even order}

In this case there are $n/2+1$ number of terms in $f_n$ and $n/2$ number
of terms in $g_n$. In this case, one of the solutions can be written in
the form (note $n$ is an even integer)
\bea\label{n9}
&&f_n = \sum_{k=1}^{(n/2+1)} A_k (\dn[\beta(n+c_2),m])^{2(k-1)}\,,
\nonumber \\  
&&g_n = \sqrt{m} \sn[\beta(n+c_2),m] \dn[\beta(n+c_2),m] 
\sum_{k=1}^{(n/2)} B_k (\dn[\beta(n+c_2),m])^{2(k-1)}\,.
\eea
For this solution, $f_n,g_n$ and hence $u_n,v_n$ satisfy the boundary
condition (\ref{m17a}). 

Another solution is given by
\bea\label{n9a}
&&f_n = \sum_{k=1}^{n/2+1} A_k (\cn[\beta(n+c_2),m])^{2(k-1)}\,,
\nonumber \\  
&&g_n = \sn[\beta(n+c_2),m] \cn[\beta(n+c_2),m] \sum_{k=1}^{n/2} B_k 
(\cn[\beta(n+c_2),m])^{2(k-1)}\,.
\eea
For this solution, $f_n,g_n$ and hence $u_n,v_n$ satisfy the boundary
condition (\ref{m19}). 

In the limit $m=1$, both the solutions (\ref{n9}) and (\ref{n9a}) 
go over to the hyperbolic solution
\bea\label{n9b}
&&f_n = \sum_{k=1}^{n/2+1} A_k (\sech [\beta(n+c_2)])^{2(k-1)}\,,
\nonumber \\  
&&g_n = \tanh[\beta(n+c_2)] \sech[\beta(n+c_2)] \sum_{k=1}^{n/2} B_k 
(\sech [\beta(n+c_2)])^{2(k-1)}\,.
\eea

Further, in the limit $m=0$, the solution (\ref{n9a}) goes over to the
trigonometric solution
\bea\label{n9aa}
&&f_n = \sum_{k=1}^{n/2+1} A_k (\cos[\beta(n+c_2)])^{2(k-1)}\,,
\nonumber \\  
&&g_n = \sin[\beta(n+c_2)] \cos[\beta(n+c_2)] \sum_{k=1}^{n/2} B_k 
(\cos[\beta(n+c_2)])^{2(k-1)}\,.
\eea

Note that these four solutions are valid in case $\mu_1,\mu_2 <0$.  
However, if $\mu_1$ and $\mu_2$ have opposite signs, say $\mu_1 <0$, 
$\mu_2 >0$, then the solution is given by
\bea\label{n9c}
&&f_n = \sum_{k=1}^{n/2+1} \frac{A_k}{(\dn[\beta(n+c_2),m])^{2(k-1)}}\,,
\nonumber \\  
&&g_n = \sqrt{m} \sn[\beta(n+c_2),m] \sum_{k=2}^{n/2+1} 
\frac{B_k}{(\dn[\beta(n+c_2),m])^{2(k-1)}}\,.
\eea
For this solution, $f_n,g_n$ and hence $u_n,v_n$ satisfy the boundary
condition (\ref{m17a}). In the limit $m=1$, this solution goes over to the 
hyperbolic solution  
\bea\label{n9d}
&&f_n = \sum_{k=1}^{n/2+1} A_k (\cosh [\beta(n+c_2)])^{2(k-1)}\,,
\nonumber \\  
&&g_n = \tanh[\beta(n+c_2)] \sum_{k=2}^{n/2+1} B_k 
(\cosh [\beta(n+c_2)])^{2(k-1)}\,.
\eea

On substituting any of the expressions for $f_n,g_n$ as given by 
Eqs. (\ref{n9}) to (\ref{n9d}) in Eq. (\ref{m13}) we again obtain $n+1$
equations which determine the $n+1$ parameters $A_k,B_k$. Some of these
relations are
\bea\label{n10}
&&A_{n/2+1}^2 = B_{n/2}^2\,, \nonumber \\
&&2A_{n/2+1} A_{n/2}+B_{n/2}^2 = 2B_{n/2} B_{n/2-1}\,, \nonumber \\
&&A_{n/2}^2+2A_{n/2+1} A_{n/2-1}+2B_{n/2} B_{n/2-1}= 
B_{n/2-1}^2 + 2B_{n/2} B_{n/2-2}\,, \nonumber \\ 
&&B_{n/2-1}^2+2A_{n/2+1} A_{n/2-2}+2B_{n/2} B_{n/2-2} \nonumber \\ 
&&+2A_{n/2}A_{n/2-1}= 2B_{n/2-1}B_{n/2-2} + 2B_{n/2} B_{n/2-3}\,, \nonumber \\ 
&&A_1^2 = 1\,,~~B_1^2+2A_1 A_2 = 0\,,~~ 
2A_1 A_3+A_2^2+2B_1 B_2 = B_1^2\,, \nonumber \\
&&B_2^2+2A_2 A_3+2A_1 A_4+2B_2 B_2 = 2B_1 B_2\,.
\eea
On comparing these results with the exact expressions for $n=2,4$,
we conjecture the following general results for arbitrary even $n$:
\bea\label{n11}
&&A_{n/2+1}=B_{n/2} = 2^{n-1}\,,
~~A_{n/2}=-n 2^{(n-3}\,,~~B_{n/2-1}=-(n-2)2^{(n-3)}\,, \nonumber \\
&&A_{n/2-1}=\frac{n(n-3)}{2!} 2^{(n-5)}\,,
~~B_{n/2-2}=\frac{(n-3)(n-4)}{2!} 2^{(n-5)}\,,
\nonumber \\
&&A_{n/2-2}=-\frac{n(n-4)(n-5)}{3!} 2^{(n-7)}\,,
~~B_{n/2-3}=-\frac{(n-4)(n-5)(n-6)}{3!} 2^{(n-7)}\,,
\nonumber \\
&&\sum_{i=1}^{n/2+1} A_i = 1\,,~~
\sum_{i=1}^{n/2} B_i = n\,.
\eea
Further, depending on if $n=4k+2$ or $4k+4$ we have the following results
($k=0,1,2,...$):
\be\label{n12}
A_1=-1\,, ~~~B_1=n\,,A_2=\frac{n^2}{2}\,, ~~if~~n=4k+2\,,
\ee
\be\label{n13}
A_1=+1\,, ~~~B_1=-n\,,~~A_2=-\frac{n^2}{2}\,,  ~~if~~n=4k+4\,.
\ee

In fact looking at these general results, we conjecture the following
expressions for various $A_i,B_i$.
\be\label{g3}
A_{n/2-k+1} = (-1)^{k} \frac{n(n-k-1)(n-k-2)...(n-2k+1)}
{k!} 2^{n-2k-1}\,,~~k=0,1,2,...,n/2\,,
\ee
\be\label{g4}
B_{n/2-k} = (-1)^{k} \frac{(n-k-1)(n-k-2)...(n-2k)}
{k!} 2^{n-2k-1}\,,~~k=0,1,2,...,n/2-1\,.
\ee

It can be verified that for $n=2,4$, the $A_i, B_i$ which followfrom Eqs. 
(\ref{n10}) to (\ref{n13})  agree with the values obtained independently 
by us above as well as in\cite{ak11}.
Further, for low values of (even) $n$, it can be numerically checked that
the $A_i, B_i$ which follow from Eqs. (\ref{g3}) and (\ref{g4}) indeed
satisfy the various relations, which follow by demanding the validity of
the constraint relation (\ref{m13}).

In the next section, we will prove that our conjectures hold for all
$n$. But first, using Eqs. (\ref{n13}), (\ref{g3}), and (\ref{g4}), we
predict that the following identities hold for all even $n$:
\be\label{g7}
\sum_{k=0}^{n/2}  (-1)^{k} \frac{n(n-k-1)(n-k-2)...(n-2k+1)}
{k!} 2^{n-2k-1} =1\,.
\ee
\be\label{g8}
\sum_{k=0}^{n/2-1}  (-1)^{k} \frac{(n-k-1)(n-k-2)...(n-2k)}
{k!} 2^{n-2k-1} = n\,.
\ee
These identities will naturally lead to the proof of our conjectures. 

As mentioned earlier, since the field Eqs. (\ref{m11}) and (\ref{m12})
are invariant under
$f_n \rightarrow \pm f_n,g_n \rightarrow \pm g_n$, hence one can trivially
write down three other solutions in both odd and even $n$ cases.

\section{Connection to Chebyshev Polynomials}

In the previous section, we have defined two families $f_n, g_n$ and
conjectured that they lead to the solutions of the constraint relation
\eqref{m13}. Here, $f_n$ is a polynomial, and $g_n$ equals a polynomial
times an extra factor, and we took their common argument to be a Jacobi
elliptic function. We now prove that these proposed functions do indeed
give rise to solutions for all $n$. To do so, we take a closer look at
the identities stated in Eqs. (\ref{g5}), (\ref{g6}), (\ref{g7}), and
(\ref{g8}). If we look at Eqs. (\ref{g5}) and (\ref{g7}) then we notice
that in both the cases, the summation goes from $k=0$ to $k=m$, where $n =
2m+1$ in Eq. (\ref{g5}) and $n = 2m$ in Eq. (\ref{g7}). In other words,
in both cases the summation goes from $k=0$ to $\lfloor n/2 \rfloor$.
Thus Eqs.
(\ref{g5}) and (\ref{g7}) change to the odd and even cases of
\be\label{4.1}
\sum_{l=0}^{\lfloor n/2 \rfloor} \binom{n-l}{l} \frac{(-1)^l
n 2^{n-2l}}{2(n-l)} = 1\,. 
\ee
\noindent It turns out that replacing the $2$ in the numerator by other
bases yields other similar identities as well, that can be numerically
verified for small $n$. To simplify the notation, define
\be\label{4.2}
f_n(x) := \sum_{l=0}^{\lfloor n/2 \rfloor} \binom{n-l}{l} \frac{(-1)^l n
(2x)^{n-2l}}{2(n-l)}\,, 
\ee
\noindent where $x \in \R$. Our identities \eqref{g5} and \eqref{g7} say
that $f_n(1) = 1$.
But before we prove this, we remark that we can also compute $f_n(x)$ for
other values of $x$ and low values of $n$. We omit writing the details
down, but doing so leads to the following (conjectured) identities:
\be\label{4.3a}
f_n(1) = 1 = \cos(2 n \pi)\,, \qquad f_n(1/2) =
 \cos(n \pi/3)\,, \qquad f_n(0) = \cos(n \pi/2)\,, 
\ee

\noindent Similarly, it is possible to propose closed-form expressions
for $f_n(-1/2)$ and $f_n(-1)$ as well. The connection between the terms
on both sides in Eq. \eqref{4.3a} is made by noting that
\be
1 = \cos(2 \pi), \qquad 1/2 = \cos(\pi/3), \qquad 0 = \cos(\pi/2).
\ee

\noindent This leads to the following result for all $x$.

\begin{theorem}
For any $\theta \in [0,\pi]$,
\be\label{4.4a}
f_n(\cos(\theta)) = \cos(n \theta)\,.
\ee
\noindent More generally, for any $x \in \R$, $f_n(x) = T_n(x)$, where
$T_n$ is the $n$th {\em Chebyshev polynomial of the first kind}.
\end{theorem}

\noindent {\bf Proof:}
We make use of an ``explicit formula" on Chebyshev polynomials:
\be\label{4.5}
T_n(x) =  \sum_{l=0}^{\lfloor n/2 \rfloor} \frac{(-1)^l n}{2}
\frac{ (n-l-1)!}{ l! (n-2l)!} (2x)^{n-2l}.
\ee

\noindent Simplifying the right-hand side, we obtain the desired result.
Note that for $x \in [-1,1]$, $x = \cos(\theta)$ for a unique $\theta \in
[0,\pi]$, and then the definition of $T_n$ proves that $f_n(\cos(\theta))
= \cos(n \theta)$. \hfill\qed\medskip

\noindent It is also easy to compute that when $x = 1, 1/2$, or $0$,
$T_n(x) = 1, \cos(n \pi/3)$, or $\cos(n \pi/2)$ respectively, as claimed
in Eq. (\ref{4.3a}).

Similarly, if we look at Eqs. (\ref{g6}) and (\ref{g8}), then we observe that
both these cases change to the odd and even cases of
\be\label{4.6}
\frac{1}{2} \sum_{l=0}^{\lfloor n/2 \rfloor} \binom{n-l-1}{l} (-1)^l
2^{n-2l} = n\,.
\ee

It remains to compute the above series. More generally, we prove:

\begin{theorem}
Let 
\be\label{4.2a}
g_n(x) = \frac{1}{2} \sum_{l=0}^{\lfloor n/2 \rfloor}
\binom{n-l-1}{l} (-1)^l (2x)^{n-2l}\,. 
\ee
Then for all $x \in \R$,
$g_n(x) = x U_{n-1}(x)$,
where $U_n(x)$ are the {\it Chebyshev polynomials of the second kind}.
\end{theorem}

\noindent {\bf Proof:}
We start from Eq. (\ref{4.5}).
On differentiating both sides, we have 
\bea\label{4.7}
T_n'(x) & = &  \sum_{l=0}^{\lfloor n/2 \rfloor}
(-1)^l \frac{n(n-l-1)!}{l! (n-2l)!} (n-2l) (2x)^{n-2l-1}\\
& = & \frac{n}{x} g_n(x).
\eea

\noindent Simplifying, we get:
\be\label{4.8}
g_n(x) = \frac{x}{n} \cdot  T'_n(x) = x U_{n-1}(x),
\ee

\noindent where the second equality is standard.
\hfill\qed\medskip

\noindent It remains to prove the original identity. But the summation is
simply $g_n(1)$, and it is well known that $U_n(1) = n+1$ for all $n$.
Hence 
\be\label{4.9}
g_n(1) =  U_{n-1}(1) = n\,. 
\ee

In fact it is now clear that for arbitrary $n$, $f_n$ and $g_n$ (as given
by Eqs. (\ref{4.2}) and (\ref{4.2a}) respectively, for any even or odd 
integer $n$) are simply
\be\label{4.10}
f_n = T_n (y) \,,~~g_n = (1-y^2)^{1/2} U_{n-1} (y)\,,
\ee
where $y$ is one of the following:
\bea
&&\dn[\beta(n+c_2),m]\,, \qquad \cn[\beta(n+c_2),m]\,, \qquad
\sech[\beta(n+c_2)]\,, \qquad \cos[\beta(n+c_2)]\,, \nonumber \\
&&\frac{1}{\dn[\beta(n+c_2),m]}\,, \qquad \cosh[\beta(n+c_2)]\,.
\eea
Note that 
\be\label{4.11}
f_n^2(y)+g_n^2(y) = T_n^2 (y) + (1-y^2) U^2_{n-1} (y) =1\,.
\ee
In other words, the solutions of the coupled equations are simply Chebyshev
polynomials with argument in terms of Jacobi elliptic functions.  
Actually, once one realizes this, then the structure of other solutions is also simplified. In particular, it is known that
\be\label{xx}
T_n(\cos(\theta)) = \cos(n\theta)\,,~~T_n(\cosh(x))=\cosh(nx)\,.
\ee

\noindent Thus, $f_n(x) = \cos(n\theta)$ or $\cosh(nx)$, which makes 
$g_n(x) = \sin(n\theta)$ or $\sinh(nx)$ respectively.
Thus, one now has a better understanding of the solutions 
(\ref{n1d}), (\ref{n9d}), (\ref{n1aa}) and (\ref{n9aa}).

\section{Other Coupled Models}

We now consider the coupled Ablowitz-Ladik (AL), coupled $\phi^6$ 
and coupled $\phi^4$
models as discussed in our previous publication \cite{ak11} 
and show that all these models
also admit solutions in terms of Lam\'e polynomials of arbitrary order.

\subsection{Solutions of a Coupled AL Model}

As shown in our previous publication \cite{ak11}, in the special case 
when $\nu_1=2\mu_1$ and $\nu_2=2\mu_2$, the coupled
Salerno model as given by Eqs. (\ref{m8}) and (\ref{m9}) 
reduces to the coupled AL model with the field equations
\be\label{m11z}
idu_n/dt +[u_{n+1}+u_{n-1}-2u_n]+(\mu_1 \mid u_n \mid^2+\mu_2\mid
v_n \mid^2)[u_{n+1}+u_{n-1}] =0\,,
\ee
\be\label{m12z}
idv_n/dt +\left[v_{n+1}+v_{n-1}-\frac{2\mu_2}{\mu_1}v_n\right]
+(\mu_1 \mid u_n \mid^2+\mu_2\mid
v_n \mid^2)[v_{n+1}+v_{n-1}] =0\,.
\ee

It is then clear that all the solutions of coupled Salerno model in terms
of Lam\'e polynomials of arbitrary order, are 
automatically the solutions of the coupled AL model and further in this case,
$\omega_1=2,\omega_2=\frac{2\mu_1}{\mu_2}$.

\subsection{Solutions of a Coupled Discrete $\phi^6$ Model} 

The field equations of the coupled discrete $\phi^6$ model discussed in our 
recent paper \cite{ak11} are
 
\be\label{4.3}
   \frac{1}{h^2} (\phi_{n+1}+\phi_{n-1}-2\phi_n)
   =a_1\phi_n-b_1 \phi_n^3+d \psi_n^2 \phi_n+[c_1 \phi_n^4+e\phi_n^2 \psi_n^2
+f\psi_n^4][\phi_{n+1}+\phi_{n-1}]\,, 
\ee
\be\label{4.4}
   \frac{1}{h^2} (\psi_{n+1}+\psi_{n-1}-2\psi_n)
   =a_2\psi_n- b_2 \psi_n^3+d \phi_n^2 \psi_n+[c_2 \psi_n^4+\frac{e}{2}\phi_n^4
+2f \phi_n^2 \psi_n^2][\psi_{n+1}+\psi_{n-1}]\,.
\ee

As shown in \cite{ak11}, solutions to these coupled equations in terms of
Lam\'e polynomials of order one and two are obtained in case 
\be\label{4.16az}
\phi_n^2+\psi_n^2 = \sqrt{\frac{1}{c_1h^2}}\,,
\ee 
and further if
\be\label{4.16a}
c_1=c_2=f=\frac{e}{2}\,,~~b_1=b_2=-d\,,~~a_1=a_2\,,~~c_1 h^2 b^2 =1\,,
~~a_1+\frac{2}{h^2}=\frac{b_1}{\sqrt{h^2c_1}}\, . 
\ee
It is then clear that the solutions (\ref{n1}) to (\ref{n1b}) and (\ref{n9}) 
to (\ref{n9b}) in terms of Lam\'e polynomials of 
arbitrary order obtained in the case of the coupled Salerno model will also be 
the solutions of this coupled model.  

\subsection{Solutions for a Coupled Discrete $\phi^4$ Model} 

In our recent publication \cite{ak11} we considered the following coupled
$\phi^4$ model
\be\label{5.1}
   \frac{1}{h^2} (\phi_{n+1}+\phi_{n-1}-2\phi_n)
   -2\alpha_1\phi_n-[2\beta_1\phi_n^2+\gamma \psi_n^2]
   [\phi_{n+1}+\phi_{n-1}]=0\,, 
\ee
\be\label{5.2}
   \frac{1}{h^2} (\psi_{n+1}+\psi_{n-1}-2\psi_n)
   -2\alpha_2\psi_n-[2\beta_2\psi_n^2+\gamma \phi_n^2]
   [\psi_{n+1}+\psi_{n-1}]=0\,,
\ee
and showed that solutions to these coupled equations can be obtained in terms 
of Lam\'e polynomials of order one and two provided
\be\label{5.1z}
\phi_n^2+\psi_n^2 = \frac{1}{2\beta_1 h^2}\,,
\ee 
and further if
\be\label{5.3}
2\beta_1=2\beta_2=\gamma\,,~~\alpha_1=\alpha_2=-\frac{1}{h^2}\,.
\ee
It is then clear that the solutions (\ref{n1}) to (\ref{n1aa}) and (\ref{n9}) 
to (\ref{n9aa}) in terms of Lam\'e polynomials (or corresponding hyperbolic
and trigonometric polynomials) of 
arbitrary order obtained in the case of the coupled Salerno model will also be 
the solutions of this coupled model.  

\section{Summary}

In this paper we have shown that for a number of {\it coupled discrete}
models, e.g., coupled Salerno, coupled Ablowitz-Ladik,  coupled $\phi^6$,
coupled $\phi^4$, there are solutions in terms of Lam\'e polynomials of
arbitrary order while the uncoupled equations do not admit solutions in
terms of Lam\'e polynomials of order two and higher. In particular, 
we showed that the Lam\'e polynomials can be reexpressed as Chebyshev 
polynomials of the relevant Jacobi elliptic function.  Many of these solutions 
are relevant to physical contexts ranging from ferroelectric materials 
\cite{aubry,abel,jcp} to multiferroics \cite{kimura,fiebig,curnoe} to a variety 
of models in field theory \cite{das,raja} in addition to various discrete contexts 
\cite{cuevas, rothos, panosbook}.  

It is important to emphasize that the stability of the various solutions found 
here remains an open issue to be explored numerically, particularly since 
the soliton solutions obtained above are of arbitrary width.  Besides, a 
comparative study of the scattering of solitons of different  discrete models 
is an important issue with these static solutions boosted with a certain velocity. 
Similarly, the Peierls-Nabarro (discreteness) barrier for the solutions remains 
to be explored. However, since all our solutions have discrete translational 
invariance (i.e., they are valid for arbitrary $c_2$), it is likely that for all our 
solutions the Peierls-Nabarro barrier may be zero.  However, this issue 
needs to be explored carefully and we intend to do so in the near future.
   
\section{Acknowledgment}
This work was supported in part by the U.S. Department of Energy.

\end{document}